\documentclass[12pt,preprint]{aastex}

\def\lessim{\mathrel{\hbox{\rlap{\hbox{\lower4pt\hbox{$\sim$}}}\hbox{$<$}}}}
\def\grtsim{\mathrel{\hbox{\rlap{\hbox{\lower4pt\hbox{$\sim$}}}\hbox{$>$}}}}

\shorttitle{Orbital Period of V368 Aquilae}
\shortauthors{Marin \& Shafter}

\begin{document}


\title{The Orbital Period of V368 Aquilae (Nova Aquilae 1936 No. 2)}


\author{E. Marin and A. W. Shafter}
\affil{Department of Astronomy and Mount Laguna Observatory\\
     San Diego State University\\
    San Diego, CA 92182}
\email{emarin@sciences.sdsu.edu,aws@nova.sdsu.edu}




\begin{abstract}

We report observations of the eclipsing classical nova
V368~Aql (Nova Aql 1936 No.~2).  These
data reveal that the orbital period previously published by
Diaz \& Bruch is an alias of the true orbital period.
A total of 14 eclipses (12 complete and 2 partial)
over 25 nights of observation have established that
the orbital period of V368~Aql is $0.6905093(1)$~d (16.57~hr),
which is roughly twice the previously published period.
With its revised orbital period, V368~Aql now joins other nova systems
with periods in excess of 0.5 day that dominate the long
end of the orbital period distribution of cataclysmic variables.

\end{abstract}





\section{Introduction}

V368~Aql (Nova Aquilae 1936 No. 2) was discovered at Kvistaberg Observatory
on 7 October 1936 by Taam (1936) as an optical transient of
roughly seventh magnitude.
The outburst light curve
based on observations by several observers (see Payne-Gaposchkin 1957;
Klemola 1968), 
showed that
maximum light occurred approximately two weeks earlier
on 1936 September 25 at $m_{pg}\simeq6.6$. The nova faded relatively quickly,
dropping by three magnitudes in $\sim42$~days ($t_3=42$~d), making V368~Aql
a moderately fast nova.
Spectroscopic observations made $\sim$25 days after maximum revealed
an absorption-line system consisting of H, Ca~II, He~II $\lambda 4686$,
N~III, and probably He~I and Fe~II, along with
strong, broad ($\sim 2000$~km~s$^{-1}$ Balmer emission (Wyse 1936).
Based on these data the spectroscopic class of V368~Aql is ambiguous,
but the probable detection of Fe~II, and the moderate width of the
Balmer lines suggest that the object was likely a member of either
the ``Fe~II" or ``hybrid" classes of novae as defined by Williams (1992).

Aside from its identification at minimum light by Klemola (1968),
V368~Aql slipped into relative obscurity until time-resolved
photometric observations by
Diaz \& Bruch (1994) revealed shallow eclipses in the light curve.
A total of just two eclipses were observed
during their 11 nights of observation
in 1990 May, and 1993 July and August.
Based on a periodogram analysis of their
entire dataset Diaz \& Bruch concluded that
the orbital period of V368~Aql was likely to be $0.34521\pm0.00015$~d
(8.29~hr).

As part of a program to study the eclipse morphologies of classical
novae, we obtained a sequence of time-resolved observations of V368~Aql
on the nights of 2005 September 3, 4, and 5 UT.
To our surprise, although the night of September 4 revealed an eclipse,
none were observed on the adjacent nights at the times expected for
an orbital period of 8.29~hr.
In order to establish the true orbital period and to
better understand the eclipse morphology,
we undertook a more extensive program of
time-resolved, multi-color CCD photometry of the eclipses of
V368~Aql. A preliminary report, establishing that
the orbital period of V368~Aql is actually twice that
originally reported by Diaz \& Bruch (1994), has been
posted in Shafter et al. (2008).
Here we provide a more detailed discussion of our
findings, including additional eclipse observations that have
enabled us to improve the precision of the revised orbital period.

\section{Observations}

A finding chart for V368~Aql based on one of our $B$-band images
from 2005 September 4 is presented in Figure~1.
Observations of V368~Aql
were carried out during 8 nights in 2005 September
and October, and 17 nights in 2008 June through August using
the Mount Laguna Observatory 1~m reflector.
On each night a series of exposures (typically 60~s)
were taken through one of the
Johnson-Cousins
$B$, $V$, $R$, $I$ filters (see Bessel 1990),
and imaged on a Loral $2048^2$ CCD.
To increase the time-sampling efficiency, only a $400\times400$
subsection of the
full array was read out. The subsection was chosen to include
V368~Aql and
several relatively bright nearby stars to be used as comparison objects for
differential photometry.
Twilight flat images of the same size were taken
and averaged for each run.

The data were processed in a standard fashion (bias subtraction and
flat-fielding) using IRAF.\footnote{
IRAF (Image Reduction and Analysis Facility) is distributed by the
National Optical Astronomy Observatories, which are operated by AURA, Inc.,
under cooperative agreement with the National Science Foundation.}
The individual images were subsequently
aligned to a common coordinate system and
instrumental magnitudes for V368~Aql
and two nearby comparison stars were then determined using the
{\it IRAF\/} APPHOT package.
Variations in atmospheric transparency were removed to first order by dividing
the flux of V368~Aql by that of one of two
nearby comparison stars marked C1 and C2 in Figure~1.
The differential light curves were then placed on an
absolute scale by calibration of the comparison stars against the
standard stars in Landolt (1992). Our calibration
reveals that stars C1 and C2 are characterized by
$B=15.49\pm0.10$, $V=14.71\pm0.10$, $R=14.10\pm0.10$, $I=13.76\pm0.10$, and
$B=16.14\pm0.10$, $V=14.79\pm0.10$, $R=13.60\pm0.10$, $I=13.11\pm0.10$,
respectively, where the errors quoted include the uncertainty in the mean
extinction coefficients used in the calibration.
The calibrated light curves of V368~Aql from our initial run in early
September 2005 are shown in Figure~2, while
the eclipse light curves from 2008 are segregated by color
and displayed in Figures~3--6.
A summary of all observations is presented in Table~1.

The eclipse light curves are quite variable with the out-of-eclipse
light levels in a given bandpass
varying by up to 0.5 magnitudes on time scales from weeks to months.
The eclipse depths for a given color are are also variable,
but at reduced level. The fact that
the eclipse depths are less variable than the out-of-eclipse
light levels suggests that most of the variability comes from
un-eclipsed regions of the system
(i.e. the outer accretion disk, the secondary star, or both.).
Not unexpectedly, the
eclipse depth is greatest in the $B$ band, where the light
level drops $\sim0.75$~mag at mid eclipse, indicating that
roughly half of the total system light is eclipsed.
At longer wavelengths the eclipse depth decreases
as the contribution from the secondary star
to the total system light begins to dominate. This increased contribution
manifests itself through the emergence of ellipsoidal variations,
which are seen most clearly in the $R$-band light curves.

\section{The Orbital Period of V368~Aql}

During our 25 nights of observation,
a total of 14 eclipses (12 complete and two partial) of V368~Aql
were observed. 
Eclipse timings for the 12 complete eclipses were measured
by fitting a parabola to the lower half of the eclipse profile
and taking the vertex as the time of mid-eclipse.
The resulting timings, plus the two available from Diaz \& Bruch (1994)
are summarized in Table~2.

Although all of the timings can be fit with the 8.29~hr period
found by Diaz \& Bruch (1994), as mentioned earlier,
their period also predicts eclipses at times when none are
observed to occur (e.g., see Fig. 2).
Thus, the true orbital period must be
an integer multiple of the 8.29~hr period. Partial eclipse observations on
consecutive nights with an ingress
observed just before sunrise on 2008 June 30 UT
followed by an egress observed just after sunset on 2008 July 1 UT (see Fig. 3)
resolved the cycle count ambiguity and
established that the true orbital period was (roughly) twice
the P=8.29~hr period.
A linear least-squares fit of all the mid-eclipse times
yields the following revised ephemeris for V368~Aql:

\begin{equation}
T_{\mathrm{mid-eclipse}} = {\rm JD}_{\odot}~2,449,189.5231(8)+0.6905093(1)~E.
\end{equation}

Residuals of the individual eclipse timings with respect to this ephemeris
are also given in Table~2, and are plotted as a function of cycle number
in Figure~7. There is no evidence for any second-order trend that might
suggest a change in period over the interval spanned by the observations.

\section{Discussion}

Our observations have revealed that the orbital period of
V368~Aql is roughly twice that previously accepted making it
one of the longest periods known among the cataclysmic variables (CVs).
The orbital period of V368~Aql is compared with the periods of other
classical novae and with CVs in general in Figure~8. Of 741 CVs with
known orbital period (Ritter \& Kolb 2003),
only 16 have periods longer than V368~Aql, with
eight of those also being nova systems.

It is well known that the orbital period distribution of novae differs
from that of CVs generally. The most striking difference is the relative
paucity of nova systems below the period gap (Log~$P_{orb}\lessim-1.0$).
This difference can be explained by the fact that systems
below the gap, where the accretion
rates driven by gravitational radiation are low,
require a higher envelope mass
to be accreted before a thermonuclear runaway (TNR) is triggered
(e.g. Townsley \& Bildsten 2005). The envelope mass required
to trigger a TNR increases
sharply with decreasing white dwarf mass.
The necessity for the larger envelope
mass coupled with the low accretion rate
requires the system to accrete for a relatively long interval between
consecutive nova outbursts, resulting in a low observed nova
rate among the short period CVs.

It is less well known that the orbital period distributions
for novae and CVs in general differs at long orbital periods as well.
For example, it is interesting to note
that although among CVs with known orbital
period novae make up just a little more than 10\% of the total, among
systems with orbital periods of more than 0.5~d, novae make up roughly
half (12 out of 25) of the total. It is likely that a combination of generally
higher mass accretion rates and higher white dwarf masses (required
for stable mass transfer) among the longest
period systems is responsible for the increased fraction of nova systems
in this period regime.

Table~3 summarizes the known properties of the 12 classical novae with
orbital periods in excess of 0.5~d. Of the 8 novae with information
on their spectroscopic class available, roughly half either belong to
the He/N or hybrid classes. This contrasts with novae at all
orbital periods where the He/N and Hybrid classes combined make
up a total of only $\sim20$\% of the total (Shafter 2007).
In addition two of the 12 long period novae are recurrent.
The relatively high percentage of He/N, hybrid, and recurrent novae among
the long period systems is consistent with the expectation that
these systems contain on average higher mass white dwarfs accreting
at higher rates.

\section{Conclusions}

We have found that the previously published orbital period
for V368~Aql ($P=0.345$~d, Diaz \& Bruch 1994)
was an alias of the true orbital period. A total of 14 eclipses
(2 partial) over 17 nights of observation have established that
the orbital period of V368~Aql is actually $0.6905093(1)$~d
(16.57~hr). The revised orbital period for V368~Aql places the
system among the other long-period nova systems that dominate
the statistics at the upper end of the CV period distribution.
Future work on V368~Aql should include a radial velocity
study and an effort to model of the eclipse profiles
in order to constrain system parameters, particularly the
white dwarf mass and accretion rate.

\acknowledgments

We thank S. R. Warren for assistance with the observations from
September 2005.
This work was supported in part by NSF grant AST-0607682 (AWS).



\clearpage

\begin{deluxetable}{cccccc}
\tablenum{1}
\tablewidth{0pt} 
\tablecolumns{6}
\tablecaption{Summary of Observations}
\tablehead{\colhead{} 					&
           \colhead{UT Time} 				& 
	   \colhead{Time Resolution\tablenotemark{a}} 	&
	   \colhead{Number of} 				&
           \colhead{}                                   & 
           \colhead{}                                  \\ 
	   \colhead{UT Date} 				& 
	   \colhead{(start of observations)} 		&
	   \colhead{(sec)} 				& 
	   \colhead{Exposures} 				& 
	   \colhead{Filter}			        & 
	   \colhead{Eclipse?}			       }
\startdata
2005 Sep 03 &04:09:10 &122.4 &90 &B & N\\
2005 Sep 04 &03:14:00 &122.4 &120 &B & Y\\
2005 Sep 05 &05:13:30 &62.06 &160 &V & N\\
2005 Sep 29 &04:05:40 &62.06 &100 &V & N\\
2005 Sep 30 &03:09:00 &62.06 &173 &V & N\\
2005 Oct 04 &02:50:00 &62.06 &199 &V & N\\
2005 Oct 05 &03:10:35 &62.06 &80 &V & N\\
2005 Oct 06 &02:45:00 &122.4 &92 &B & N\\
2008 Jun 29 &05:06:00 &62.06 &377 &B & N\\
2008 Jun 30 &04:39:11 &62.06 &398 &B & Y\\
2008 Jul 01 &04:38:00 &62.06 &414 &B & Y\\
2008 Jul 02 &04:28:00 &62.06 &155 &B & N\\
2008 Jul 05 &04:54:00 &62.06 &326 &B & Y\\
2008 Jul 07 &04:28:00 &62.06 &410 &B & Y\\
2008 Jul 12 &04:12:00 &62.06 &198 &V & Y\\
2008 Jul 14 &04:27:02 &62.06 &424 &R & Y\\
2008 Jul 16 &05:45:00 &62.06 &359 &I & Y\\
2008 Jul 23 &04:14:00 &62.06 &435 &R & Y\\
2008 Jul 25 &05:16:00 &62.06 &369 &V & Y\\
2008 Jul 27 &04:12:00 &62.06 &421 &I & Y\\
2008 Aug 01 &04:03:00 &62.06 &419 &B & Y\\
2008 Aug 02 &04:20:00 &62.06 &353 &R & N\\
2008 Aug 03 &03:58:00 &62.06 &445 &R & Y\\
2008 Aug 05 &03:40:00 &62.06 &420 &V & Y\\
2008 Aug 06 &03:44:00 &62.06 &417 &V & N\\
\enddata
\tablenotetext{a}{Mean time interval between exposures (integration time plus 
readout time)}
\end{deluxetable}

\clearpage

\begin{deluxetable}{cccr}
\tablenum{2}
\tablewidth{0pt}
\tablecolumns{4}
\tablecaption{Eclipse Timings}
\tablehead{\colhead{HJD (mid-eclipse)}                  &
           \colhead{Cycle Number}                       &
           \colhead{}                                   &
           \colhead{$O-C$}                             \\
           \colhead{(2,400,000+)}                       &
           \colhead{(E)}                                &
           \colhead{Filter}                             &
           \colhead{($\times 10^-3$ day)}                      }
\startdata
49189.522 &0    &R\tablenotemark{1} &$-1.089268$\\
49227.502 &55   &R\tablenotemark{1} &0.900800\\
53617.760 &6413 &B &0.952550\\
54652.833 &7912 &B &0.554559\\
54654.905 &7915 &B &1.026745\\
54659.738 &7922 &V &0.461844\\
54661.810 &7925 &R &0.934030\\
54663.880 &7928 &I &$-0.593785$\\
54670.787 &7938 &R &1.313500\\
54672.856 &7941 &V &$-1.214314$\\
54674.927 &7944 &I &$-1.742129$\\
54679.762 &7951 &B &$-0.307029$\\
54681.834 &7954 &R &0.165156\\
54683.904 &7957 &V &$-1.362659$\\
\enddata
\tablenotetext{1}{Kron-Cousins $R$ filter}
\end{deluxetable}

\clearpage

\begin{deluxetable}{lccccc}
\tablenum{3}
\tablewidth{0pt}
\tablecolumns{4}
\tablecaption{Long Orbital Period Novae}
\tablehead{& & \colhead{Orbital Period}    &
           \colhead{$t_3$}                 &
           \colhead{$V_{exp}$}             &
           \colhead{}                      \\
           \colhead{Nova}                  &
           \colhead{Type}                  &
           \colhead{(days)}                  &
           \colhead{(days)}                  &
           \colhead{(km~s$^{-1}$)}        & 
           \colhead{Spectroscopic Class}   }
\startdata
DI Lac     &  Na   &    0.5438 &  43 &  \dots & Fe II? \\
V458 Vul   &  Na   &    0.5895 &  15 &  3000  & Hy   \\
V841 Oph   &  Nb   &    0.6013 & 130 &  \dots & \dots \\
V368 Aql   &  Na   &    0.6905 &  42 &  \dots & Fe~II, hybrid? \\
V723 Cas   &  Nb   &    0.6933 &\dots& 500?   & Fe II \\
V394 CrA   &  RN   &    0.7577 &  6  &        &      \\
CP Cru     &  Na   &    0.9440 &\dots& 2000   & He/N \\
U Sco      &  RN   &    1.2300 &  7  &        & He/N? \\
X Ser      &  Nb   &    1.4800 & 170 &  \dots & \dots \\
J0305+0547 & \dots &    1.7700   &\dots&  \dots & \dots \\
GK Per     &  Na   &    1.9968 &  13 & 4000   & He/N \\
V1017 Sgr  &  Nb   &    5.7140 & 130 &  \dots & Fe II \\

\enddata
\end{deluxetable}



\clearpage

\begin{figure}
\epsscale{0.8}
\plotone{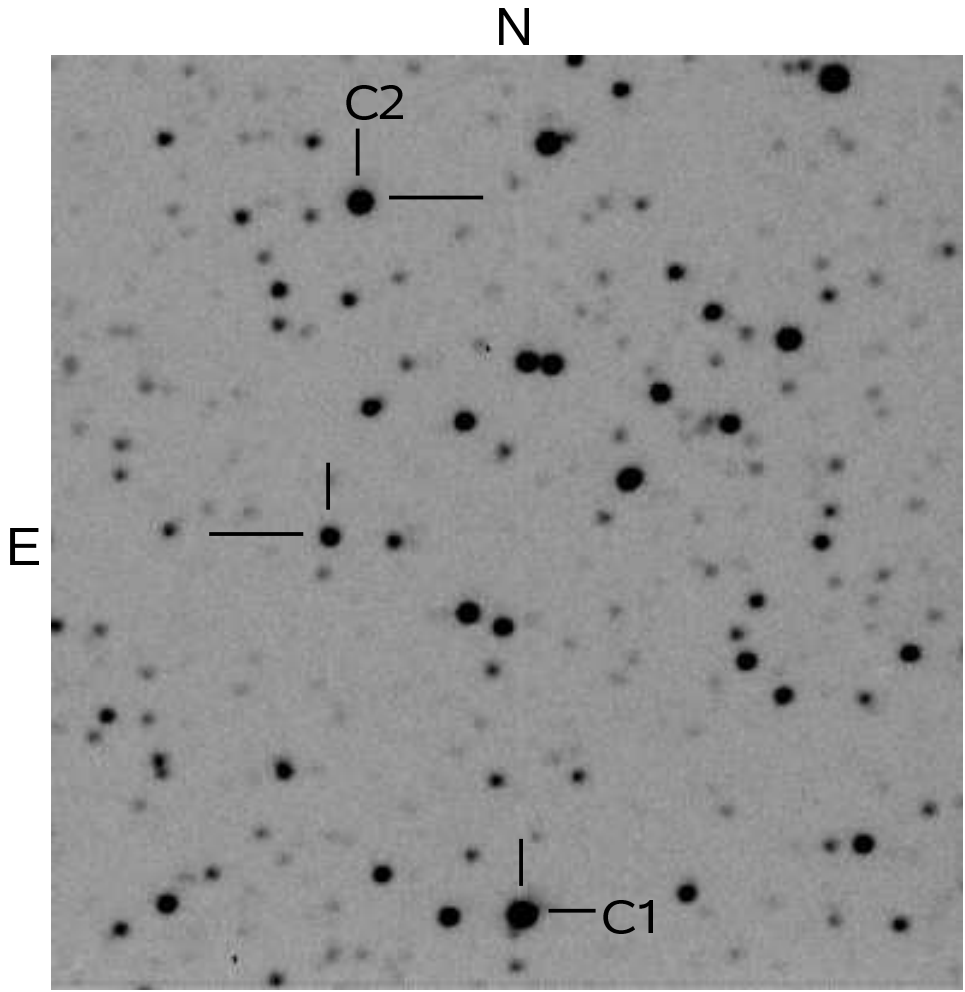}
\caption{A finding chart for V368~Aql, which is located at
         RA. = 19:26:34.45, DECL. = +07:36:13.23 (equinox 2000.0)
         based on the position given in Klemola (1968).
         The comparison stars, C1 and C2, used to calibrate the data
         are located $\sim32''$W, $\sim63''$S, and
         $\sim5''$W, $\sim56''$N of V368~Aql, respectively.}
\end{figure}

\begin{figure}
\epsscale{0.8}
\plotone{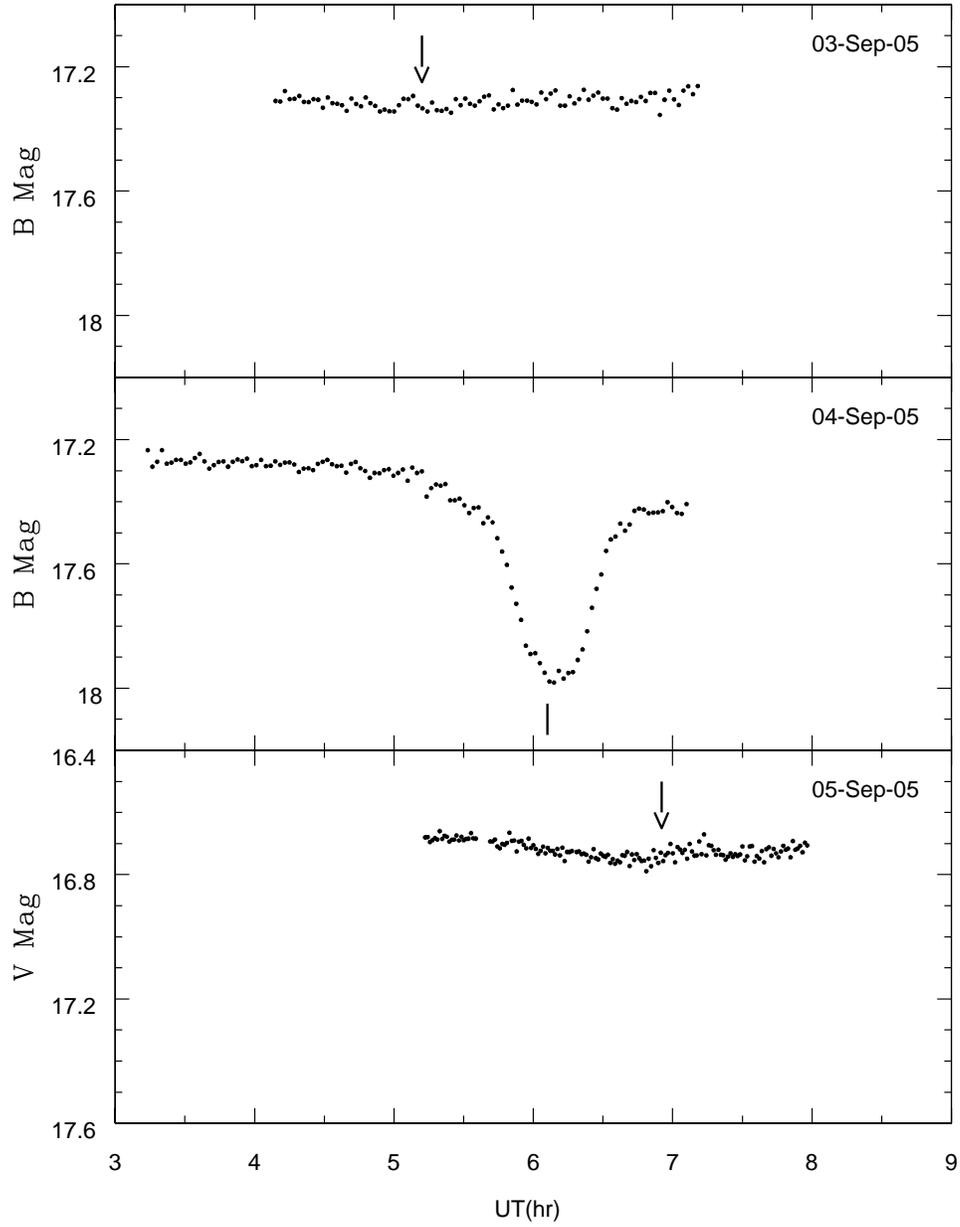}
\caption{Light curves of V368~Aql from September 2005. An eclipse
         was observed on 2005 September 4 UT, but no eclipses were observed
         on the adjacent nights at the times (indicated by arrows)
         expected assuming the 8.29~hr period of Diaz \& Bruch (1994).
         }
\end{figure}

\begin{figure}
\epsscale{0.8}
\plotone{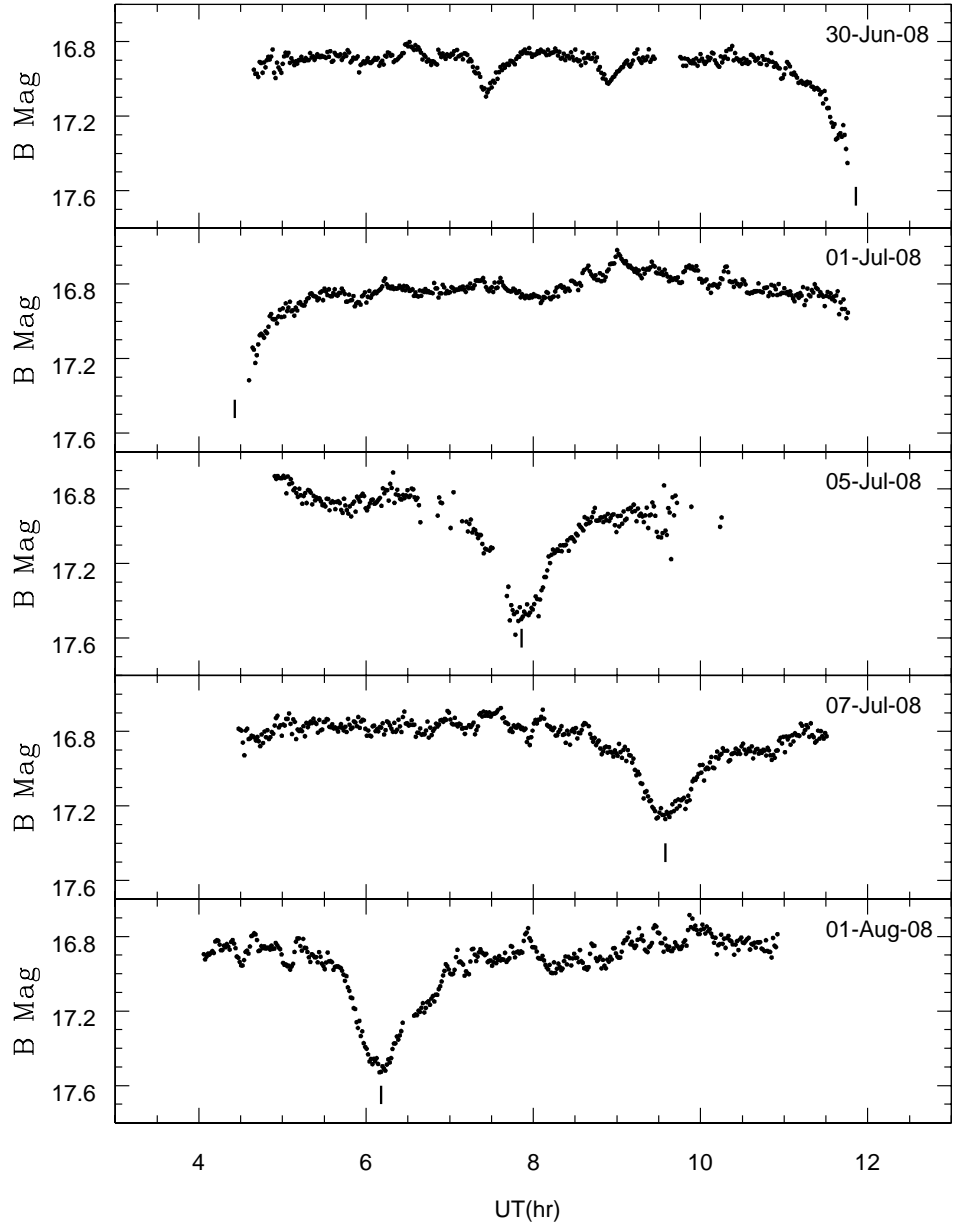}
\caption{The $B$-band light curves of V368~Aql from 2008. The eclipse
         depth is $\sim 0.75$ mag, indicating that approximately
         half of the total $B$-band light from the system is blocked
         at mid-eclipse.
         Note the partial eclipses observed on June~30 UT (ingress) and
         July~1 UT (egress).
         The tick marks under the eclipses here and in other light curves
         indicate the times of mid eclipse computed using eqn~(1).
         }
\end{figure}

\begin{figure}
\epsscale{0.8}
\plotone{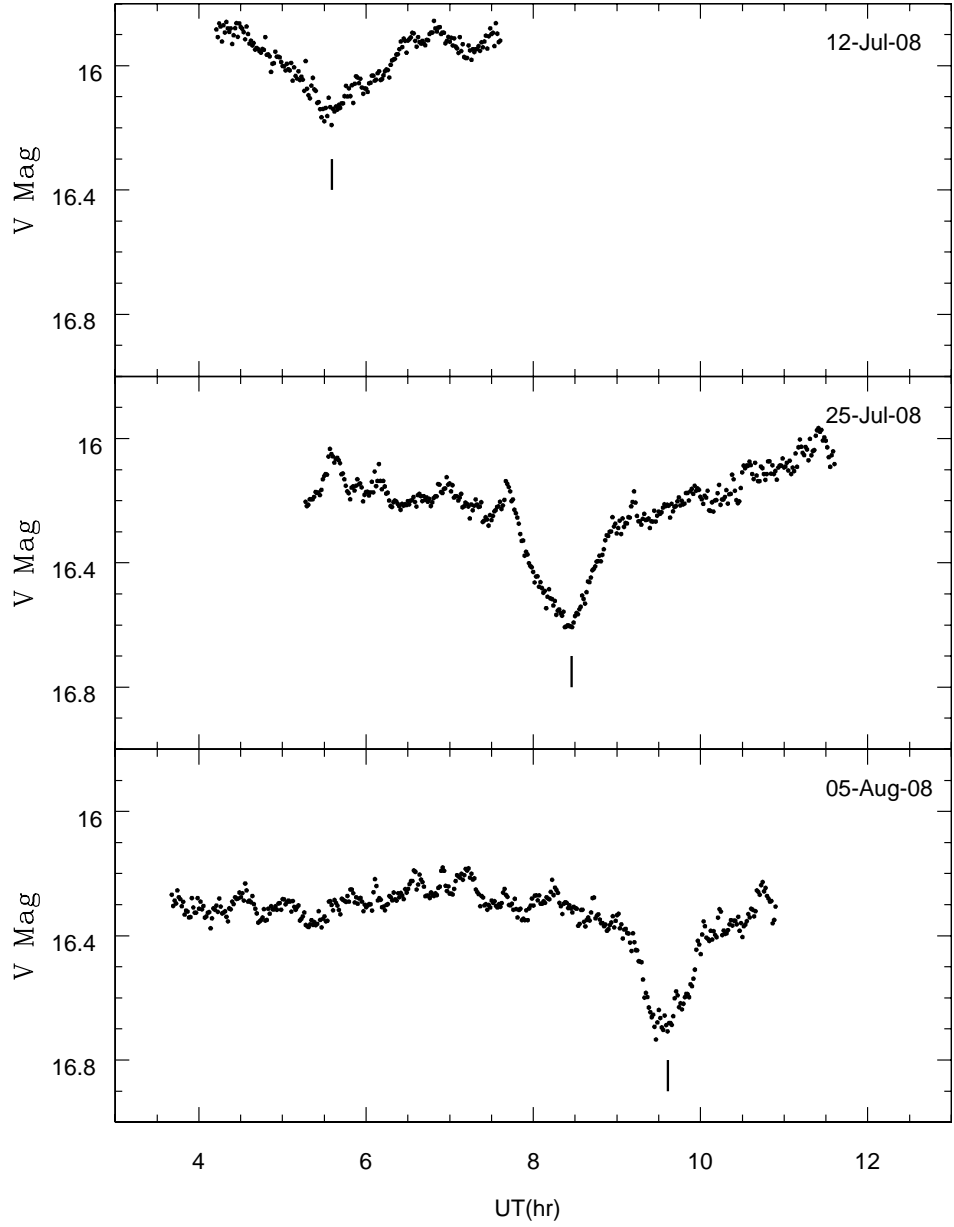}
\caption{The $V$-band eclipse light curves. Note that the out-of-eclipse
         light level was somewhat variable, with a mean eclipse
         depth in $V$-band light of $\sim 0.4$~mag.}
\end{figure}

\begin{figure}
\epsscale{0.8}
\plotone{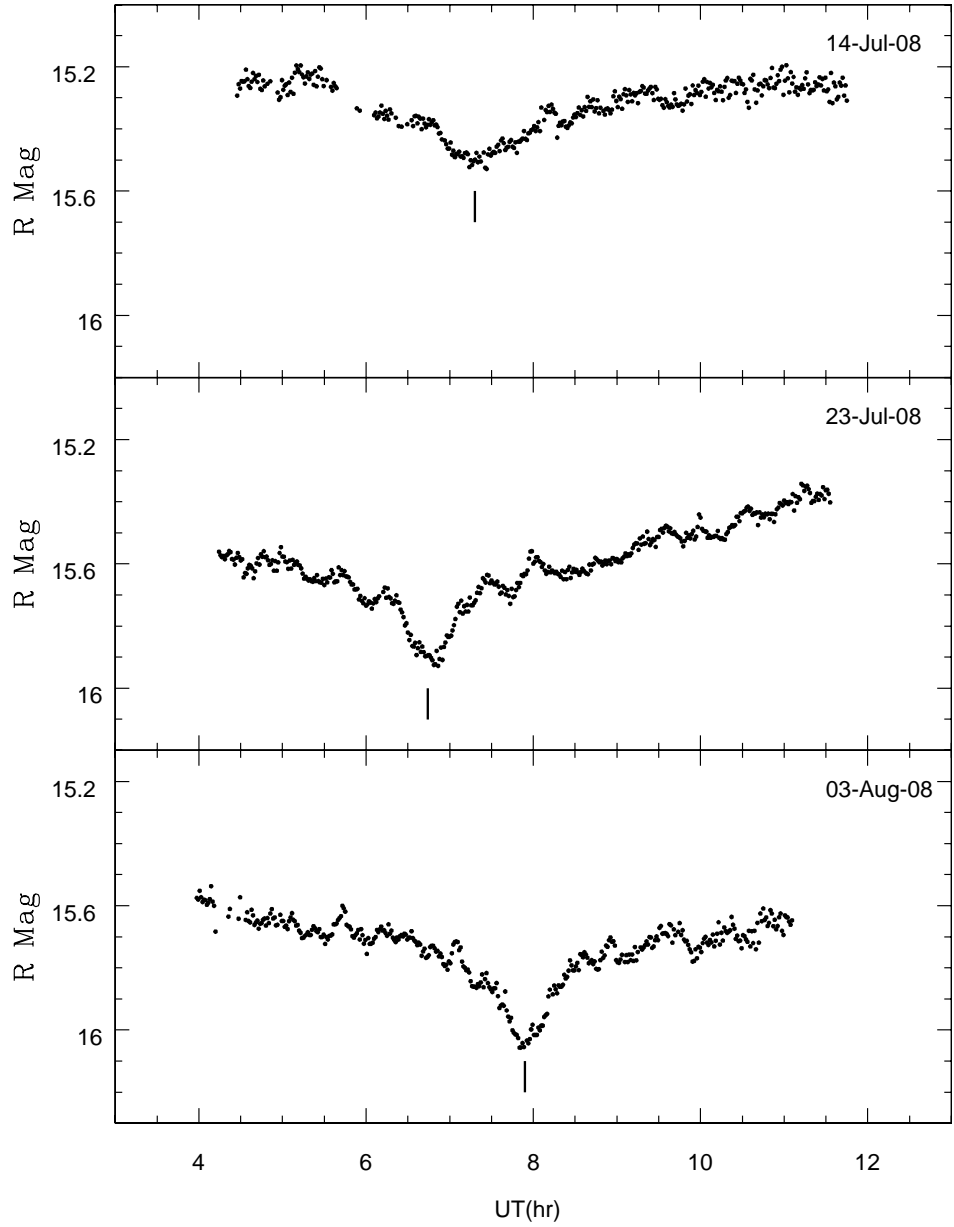}
\caption{The $R$-band eclipse light curves. As with the $V$ band data
         the out-of-eclipse light level is variable, with the mean
         eclipse depth $\sim 0.3$~mag. The variation of the out-of-eclipse
         light level on a given night results from the changing
         aspect of the Roche-lobe-filling secondary star (ellipsoidal
         variations), which dominate the light at longer wavelengths.}
\end{figure}

\begin{figure}
\epsscale{0.8}
\plotone{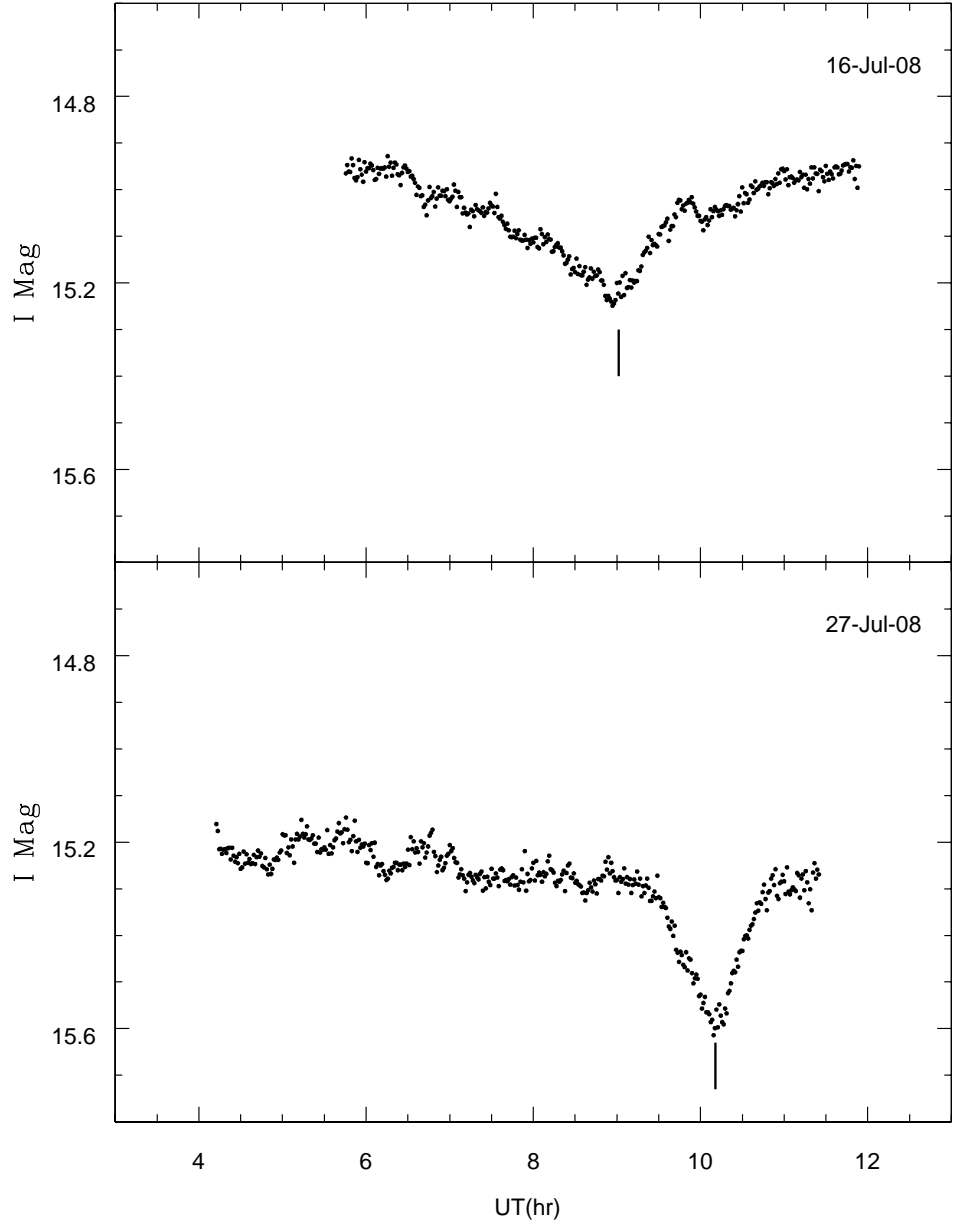}
\caption{The $I$-band eclipse light curves.
         Note again the variability in the out-of-eclipse light level.
         The mean eclipse depth ($\sim 0.3$~mag) seen here and in the
         $R$-band light curves is
         slightly less than half that seen in the $B$-band eclipses,
         and is caused by the greater contribution to the total light
         of the late-type secondary star at longer wavelengths.}
\end{figure}

\begin{figure}
\epsscale{0.8}
\plotone{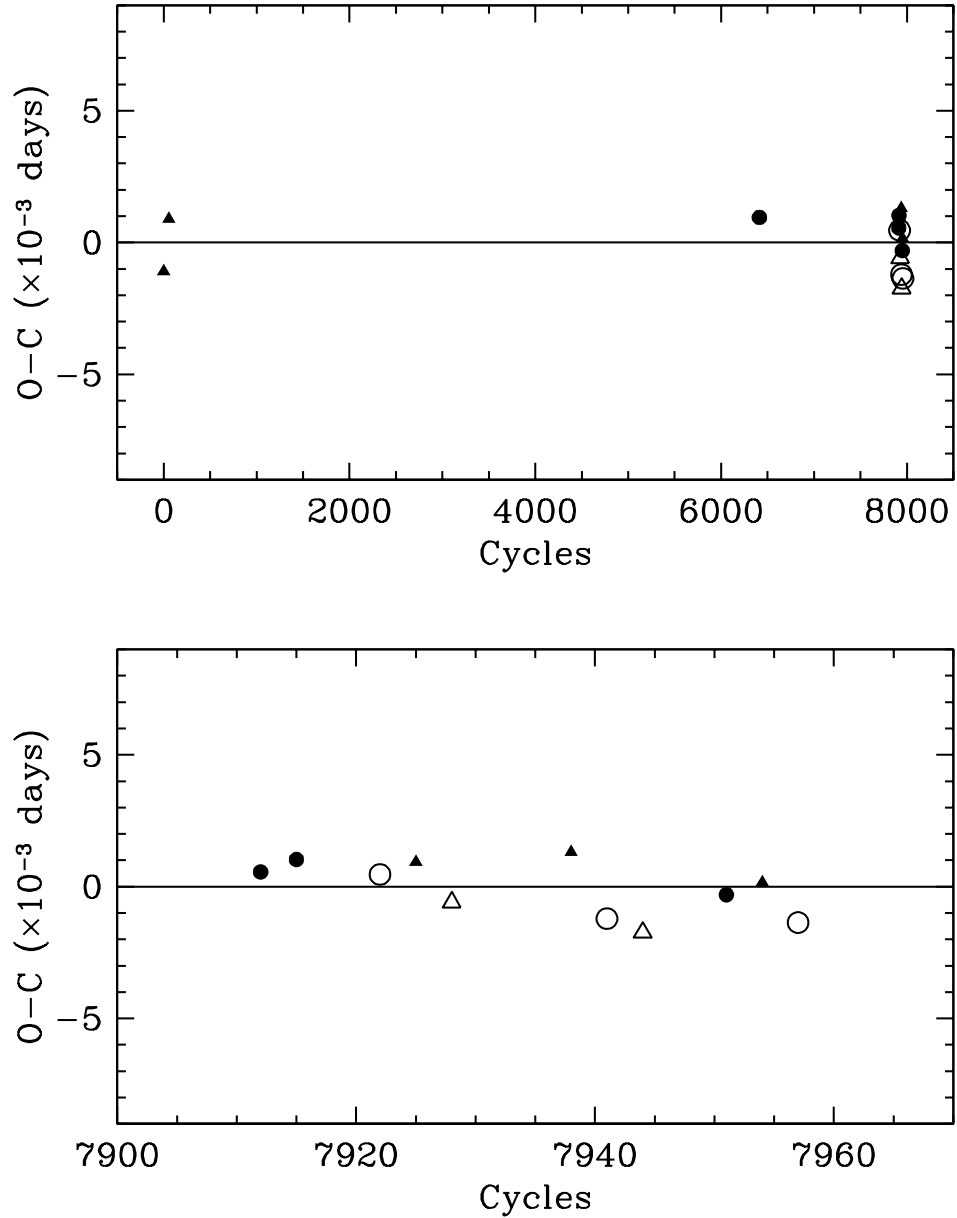}
\caption{The residuals of the observed times of eclipse with respect
         to the times calculated from eqn~(1) are plotted as a function of
         cycle number. {\it Top Panel:} All data. {\it Bottom Panel:}
         Expanded view of 2008 data. {\it Key:} $B$-band: filled circles,
         $V$-band: open circles, $R$-band: filled triangles, $I$-band:
         open triangles. There is no evidence for a period
         change over the time spanned by the observations.}
\end{figure}

\begin{figure}
\epsscale{0.8}
\plotone{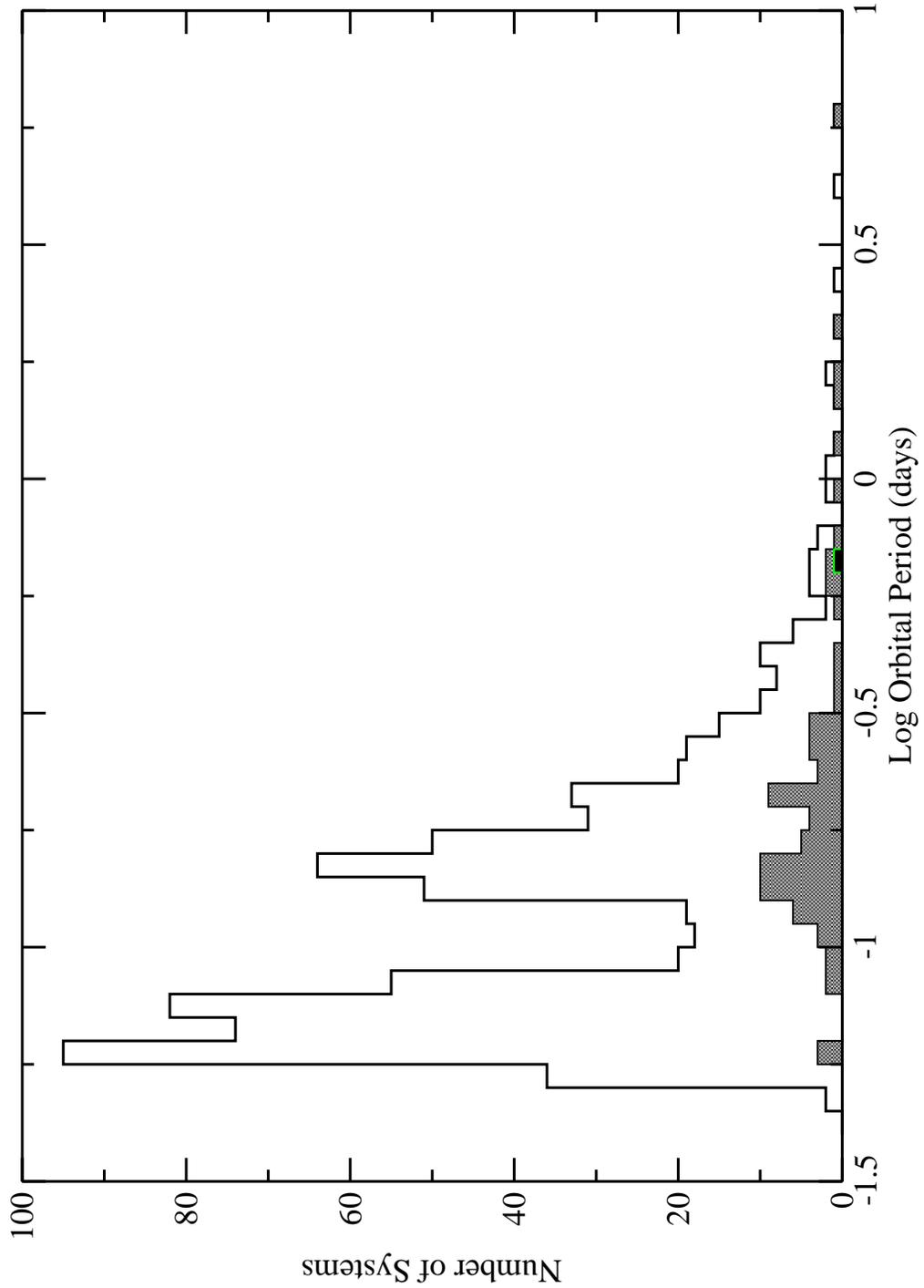}
\caption{The orbital period distribution of novae (shaded histogram)
         compared with that for all CVs with known orbital period
         (open histogram).
         V368~Aql (filled box) has the 17th longest orbital period
         known among all
         741 CVs with orbital periods longer than 0.05~d (1.2~hr).
         Of the 16 CVs with longer orbital periods half (8) are novae.
         }
\end{figure}

\end{document}